\documentclass[prb,superscriptaddress,letterpaper,10pt,twocolumn]{revtex4}
\usepackage{amsmath}
\usepackage{amssymb}
\usepackage{graphicx}
\usepackage{color}
\usepackage{epsfig}
\usepackage{dcolumn}
\usepackage{bm}
\usepackage{bm}
\usepackage{blindtext}
\usepackage{natbib}

\def\be{\begin{equation}} \def\ee{\end{equation}}
\def\bea{\begin{eqnarray}} \def\eea{\end{eqnarray}}

\begin{document}
\title{Crossovers and critical scaling in the one-dimensional transverse field Ising model}
\author{Jianda Wu}
\affiliation{Department of Physics \& Astronomy, Rice University, Houston, Texas 77005, USA}
\affiliation{Max-Planck-Institut f\"ur Physik komplexer Systeme, Dresden, 01187, Germany}
\author{Lijun Zhu}
\affiliation{Department of Physics and Astronomy, University of California, Riverside, CA 92521, USA }
\author{Qimiao Si}
\affiliation{Department of Physics \& Astronomy and Rice Center for Quantum Materials, Rice University, Houston, Texas 77005, USA}

\date{\today}
\begin{abstract}
We consider the scaling behavior of thermodynamic quantities in the one-dimensional
transverse field Ising model near its quantum critical point (QCP).
Our study has been motivated by
the question about
the thermodynamical signatures of this paradigmatic
quantum critical system and, more generally,
by the
issue of
how quantum criticality accumulates
entropy.
 We find that the crossovers
in the phase diagram of temperature and
(the non-thermal control parameter) transverse field
obey
a general scaling ansatz, and so does the
critical scaling behavior of the specific heat
and magnetic expansion coefficient.
Furthermore, the Gr\"{u}neisen
ratio diverges in a power-law way when the QCP is accessed as a function of the transverse
field at zero temperature,
which follows the prediction of quantum critical
scaling. However, at the critical field, upon decreasing the temperature, the Gr\"uneisen ratio
approaches a constant instead of
showing the expected divergence.
We are able to understand this unusual result
in terms of a peculiar form of the quantum critical scaling function for the free energy;
the contribution
to the Gr\"uneisen ratio vanishes
at the linear order in a suitable Taylor expansion of the scaling function. In spite of this
special form of the scaling
function, we show that the entropy is still maximized near the QCP, as expected from the general
scaling argument. Our results establish the telltale thermodynamic signature of a transverse-field
Ising chain, and will thus facilitate the experimental
identification of this model quantum-critical system in real materials.
\end{abstract}

\maketitle
%---------------------------------------------------------
\section{Introduction}

Quantum phase transition arises in a many-body system at zero temperature,
when the energy of its ground state is tuned by a non-thermal parameter to
a point of non-analyticity \cite{SpecialIssue2010,Sachdev2011a,Coleman2005, Si_Science2010}.
When a transition is continuous, it describes a quantum critical point (QCP).
The interplay between thermal and quantum fluctuations strongly influences
physical properties near a QCP.
\cite{Hertz1976, Chakravarty1989b, Millis1993,Ruegg2014,Schuberth2016,Wu2016b,Wu2017}.
In this context, an important thermodynamic quantity
is the Gr\"uneisen ratio \cite{Gruneisen1912} ---
the ratio of the thermal expansion coefficient to the specific heat.
Ref.~[\onlinecite{Zhu2003}] advanced a quantum critical scaling form for
the free energy, and demonstrated that the Gr\"uneisen ratio diverges.
This divergence is to be
contrasted with the case of the classical ($i.e.$ thermally-driven) critical
point, where the Gr\"uneisen ratio remains finite \cite{Gruneisen1912,Landau1968}.
The divergence of the Gr\"uneisen ratio has been experimentally
observed in heavy fermion metals \cite{Gegenwart2008}. Indeed, such a
divergence has been established as a means of diagnosing quantum criticality.
An important consequence of this divergence is that the entropy
is maximized near the QCP \cite{Zhu2003,Wu2011E}, which has recently been
directly shown experimentally in
several quantum critical
materials, including Sr$_3$Ru$_2$O$_7$ \cite{Rost2009} and
CeCu$_{6-x}$Au$_x$ \cite{Grube2017}.

\begin{figure}[t]
\begin{center}
\includegraphics[width=7.5cm]{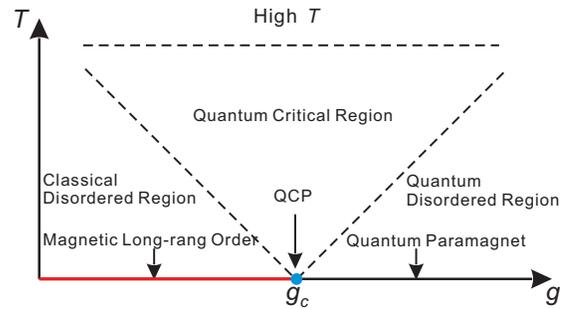}
\end{center}
\caption{The phase diagram of the 1DTFIM. $g$ denotes the transverse field
measured in unit of $J$,
with $g_c $ being the QCP.
The red solid line shows the ordered region at zero temperature.
The two tilted dash lines illustrate crossover boundaries between
different disordered regions at
nonzero temperatures.
The dashed line on the top silhouettes the crossover
to the classical region at high temperatures.} \label{fig:1DTFIMphase}
\end{figure}

In this paper we focus on the one-dimensional transverse field Ising model (1DTFIM) --
a paradigmatic system for quantum criticality
\cite{Kinross2014, Niemeijer1967,Pfeuty1970, McCoy1971,Suzuki1971, Suzuki1976, Jullien1978,Chakravarty2005}.
The 1DTFIM and its related one-dimensional models remain a topic of
considerable interest
\cite{Coldea2010,Wu2014,Perk1984,Zam1989,Fateev1994,Delfino1996,Fonseca2003,Carr2003,Perk2009}.
The 1DTFIM undergoes a continuous quantum phase transition
when the transverse field, which we measure in unit of the nearest-neighbor exchange
interaction and define as $g$ [see Eq.~(\ref{1}) below],
is tuned across its critical value, $g_c$.
Near $g_c$, the correlation length diverges as $|g - g_c|^{-1}$ ,
and the
excitation gap
%of the system
 closes as $|g - g_c|$
with the correlation length and dynamic
 critical exponents $\nu = z = 1$.
% \cite{Sachdev2011a}.
The divergence of the correlation length implies
the existence of long-range entanglement of the ground state wavefunction
near the QCP, which is a linear-superposition of
enormous possible product states of spins
in the huge $2^N$ Hilbert space with $N$ being the total number of sites
of this thermodynamic system.
This long range entanglement strongly influences the physics around the QCP,
leading to a rich $T-g$
phase diagram
shown in Fig.~\ref{fig:1DTFIMphase}, where
the crossover boundaries between the
different regimes are characterized by $T \propto |g - g_c|$ [\onlinecite{Suzuki1976}].

The hyperscaling ansatz \cite{Zhu2003} leads to general statements
about the crossover and scaling behavior of the thermodynamic quantities
near a QCP. While such an analysis is very powerful, what the ansatz does
not specify is the actual
form of the scaling functions. In order to identify
signatures that are unique to the 1DTFIM, here we systematically analyze
such crossover and scaling properties by determining the scaling functions.
We demonstrate that the entropy is indeed
maximized near the QCP of the 1DTFIM. We also show that the specific heat
and magnetic expansion coefficient [{\it c.f.} Eq.~(\ref{G1})] exhibit crossovers and scaling behavior
compatible with the hyperscaling ansatz, but with some unique features
that serve as telltale signs for the 1DTFIM.
These features can be used  to experimentally ascertain
whether the 1DTFIM is realized in candidate materials.

The specific features we have identified are as follows.
While the free energy for the 1DTFIM does obey the hyperscaling ansatz,
there is an unusual contrast between the $T$ and $g$ dependencies of the Gr\"uneisen ratio.
%The crossover scaling of its
% $T$-dependent one
% $T$-dependence
%does not follow the relation $T \propto |g - g_c|$,
%while that of its $g$-dependence does.
%The contrast becomes sharper when comparing the critical behaviors
%in the two different limits.
We
find that the Gr\"uneisen ratio
is power-law divergent when
the system is tuned through the QCP at zero temperature by varying the transverse field,
but approaches
a finite constant when the QCP is reached with decreasing temperatures.
The former respects the conventional analysis based on
the hyperscaling ansatz \cite{Zhu2003,Garst2005}.
The later, however, fails to display
the expected singular behavior.
We identify the origin of this special feature in the quantum critical region
: In the 1DTFIM,
the non-vanishing scaling term in
this part of the phase diagram ({\it c.f.}, Fig.~\ref{fig:1DTFIMphase})
begins at the second order.

The remainder of the paper is organized as follows.
 Sec.~\ref{sec:general} is devoted to a general discussion of
the critical scaling behaviors
of the Gr\"uneisen ratio near generic QCPs.
Sec.~\ref{sec:thermodynamicsin1DTFIM} specifies
the exact integral expressions for the thermodynamic quantities of the 1DTFIM.
In Sec.~\ref{sec:scalingin1DTFIM},
we determine the scaling function for the free energy of the 1DTFIM
in the classical disordered, quantum disordered and quantum critical regions;
in turn, we extract the analytic expressions that describe
the critical scaling of various thermodynamic quantities.
Sec.~\ref{sec:crossoverscalingin1DTFIM} presents the numerical results
on the crossover scaling behaviors for these thermodynamic quantities,
Sec.~\ref{sec:discussions} contains the discussions and conclusions.

\section{the Gr\"uneisen ratio near generic magnetic-field-tuned QCPs} \label{sec:general}

When the control parameter is an external magnetic field $H$,
the Gr\"uneisen ratio is defined as \cite{Gruneisen1912, Landau1968, Zhu2003}
\be
\Gamma _H  = \frac{{\alpha _H }}{{c_H }} =  - \frac{1}{T}\frac{{\left( {\partial M/\partial T} \right)_H }}{{\left( {\partial S/\partial T} \right)_H }} =  - \frac{1}{T}\frac{{\left( {\partial S/\partial H} \right)_T }}{{\left( {\partial S/\partial T} \right)_H }}
 \label{G1},
\ee
where $c_H$, $\alpha_H$, and $S$ are
the molar specific heat, magnetic expansion
coefficient, and entropy, respectively.
In addition $M$ is the magnetization per mole.
%with $V_m = V/N$ being the molar volume.
%If the control parameter is
%not linearly-coupled to pressure,
% but others
%such as an external
%magnetic field $H$,
%then the corresponding
%the parameter $p$ in
%Eq.~(\ref{G1}) should be replaced
%by the  external field $H$.
%accordingly.
%Following the scaling argument
The scaling analysis leads to the expectation that
the Gr\"uneisen ratio
%is expected to be
is generically
divergent at QCPs \cite{Zhu2003}. Since the effective dimension ($d+z$)
for the 1DTFIM is 2 ($d=1,z=1$),
the quantum criticality is expected to be non-Gaussian, and
% thus
the hyperscaling hypothesis should
work well for the model
in the thermodynamic limit. In the hyperscaling ansatz,
%if assuming
 the
critical behavior is governed by the correlation length along
the space and time dimensions, $\xi$ and $\xi_\tau$,
%then
and the
critical component of the
free energy of
a  thermodynamic system
near its critical point, $F_{cr}$, can
be generally written as \cite{Zhu2003}
\bea
\frac{{F_{cr} }}{N} & = &  - \rho _0 |r|^{\nu (d + z)} \tilde
f(\frac{T}{{T_0 |r|^{\nu z} }}) \label{F0} \\
& = & - \rho _0 \left(
{\frac{T}{{T_0 }}} \right)^{(d + z)/z} f\left( {\frac{r}{{\left(
{T/T_0 } \right)^{1/(\nu z)} }}} \right), \label{F1}
 \eea
where $\rho_0$ and $T_0$ are nonuniversal constants;
$r = (H - H_c)/H_c $ is the magnetic-field-tuning parameter
normalized by the critical field $H_c$, shifted such that
the QCP corresponds to $r = 0$ at $T=0$;
%while
and
$f(x)$
and $\tilde f(x)$ are universal scaling functions.
%Then from general scaling analysis scaling behaviors of
It then follows that the Gr\"uneisen ratio near a generic
%QCPs follow by
QCP has the following forms \cite{Zhu2003}
\bea
 & &\Gamma _{cr} (T,r = 0) = \frac{{\alpha _{cr} }}{{c_{cr} }} =  - G_T T^{ - 1/(\nu z)},   \label{G2} \\
 & &\Gamma _{cr} (T \to 0,r) =  - G_r \frac{1}{{V_m (H - H_c )}},
 \label{G3}
 \eea
where
%$r=p-p_c$
$G_T  = \{ [(d + z - 1/\nu )zf'(0)]/[d(d +
z)f(0)]\} [T_0^{1/(\nu z)} /(H_c V_m )]$ and $G_r  = \nu (d - y_0
z)/y_0 $ with $y_0$ coming from the expansion of $\tilde f(x\to 0) =
\tilde f(0)+c x^{y_0+1}$;
%. (Here the existence of
$y_0$ is introduced to
guarantee that the entropy vanishes in the zero temperature
limit.
%).
%(Note, also, that we have taken
%$r=p-p_c$.)
It is expected that $G_r \le 0 $ due to the stronger competition
in the quantum critical region between
non-commutative fields than that in the disordered region~\cite{Garst2005};
%,as it is argued in the Ref.[\onlinecite{Garst2005}].
%Indeed
this is indeed true for the 1DTFIM.
Then from Eq.~(\ref{G3}) one can observe that when the
magnetic field $H$ varies from values smaller to larger than
% the QCP,
$H_c$, the
sign of the Gr\"uneisen ratio changes from negative to positive. As a
result, from the original definition of the Gr\"uneisen ratio
(Eq.~(\ref{G1})) the entropy
at a low but nonzero temperature
generally is
maximized near
%generic QCPs at low temperatures.
the QCP.
%It shall be
We shall explicitly demonstrate
%d
that Eq.~(\ref{G3}) is correct
%in
for the 1DTFIM.
 %However, as a sharp contrast,
By contrast,
we will show that Eq.(\ref{G2}) does not hold in the 1DTFIM although
the free energy follows the form of the hyperscaling ansatz.

\section{Thermodynamic quantities in the 1DTFIM} \label{sec:thermodynamicsin1DTFIM}
The 1DTFIM with nearest neighboring interaction
%follows by \cite{Sachdev2011a},
corresponds to the following Hamiltonian,
\be H_I  =  - J\sum\limits_i {\left( {g\hat
\sigma _i^x  + \hat \sigma _i^z \hat \sigma _{i + 1}^z } \right)},
\label{1} \ee
where the spins $S_i^\alpha = \frac{1}{2}\sigma _i^\alpha$$(\alpha=x,y,z)$
with $\sigma _i^\alpha$ and $i$ denoting the Pauli matrices and the site positions, respectively.
The nearest neighbor distance is set
to 1.
We consider ferromagnetic exchange interactions, with
the nearest neighbor
coupling $J$, between the Ising ($z$-)
component of the spins, taken to be positive
$J > 0$.
In addition,
$gJ$
describes
an external magnetic field
for the transverse ($x$-) component of the spins.
As is standard, after applying
the Jordan-Wigner transformation,
\be
\left\{ \begin{array}{l}
 \sigma _i^x  = 1 - 2c_i^\dag  c_i  \\
 \sigma _i^z  =  - \prod\limits_{j < i} {\left( {1 - 2c_j^\dag  c_j } \right)} \left( {c_i  + c_i^\dag  } \right) \\
 \end{array} \right.,
\label{3}
\ee
and Bogoliubov quasi-particle transformation
$\gamma _k  = u_k c_k  - iv_k c_k^\dag$ with conditions of $u_k ^2 +
v_k ^2 =1, u_{-k}=u_k$, $v_{-k}=-v_k$, and $c_k  = \frac{1}{{\sqrt N }}\sum\limits_j {c_j e^{
- ik j } }$,
one obtains a diagonalized
Hamiltonian in the Bogoliubov quasi-particle representation,
% is obtained,
\be
H_I  = \sum\limits_k {\varepsilon _k \left( {\gamma _k^\dag  \gamma
_k  - 1/2} \right)} \label{6}
\ee
with the single-particle spectrum $\varepsilon _k  = 2J( 1 + g^2  - 2g\cos k
)^{1/2}$.
% (the nearest neighbor distance is set as 1).
%Then the
It follows that the free energy density for the 1DTFIM,
%follows by (it is reduced
%to be dimensionless
normalized by $J$, has the form \cite{Katsura1962,Pfeuty1970}
\be
f_I = \frac{F_I}{NJ} =   - \frac{1}{\beta }\left[ {\ln 2 + \frac{1}{\pi }\int_0^\pi
{dk\ln \cosh \left(\frac{\beta \varepsilon_k}{2} \right)} } \right],
%;
\label{7}
\ee
where the dimensionless temperature $t=k_B T/J$ and $\beta = 1/t$ are introduced for convenience.
And the $\varepsilon_k$ here and
in the following does not contain the factor of $J$ any more.
With the free energy density Eq.~(\ref{7}), the (dimensionless) entropy $s=-(\partial f_I/\partial t)$,
%"thermal"
magnetic
expansion coefficient $\alpha = - (\partial s /\partial g)_t$, and specific heat $c_v=t(\partial s/
\partial t)_g$ are determined as follows
\bea
s &= &\ln 2 + \nonumber\\
&&\frac{1}{\pi } \int_0^\pi {dk\left[ {\ln \cosh \left(
{\frac{\beta \varepsilon_k}{2} } \right) - \frac{\beta \varepsilon_k}{2} {\tanh \left( {\frac{\beta \varepsilon_k}{2} } \right)}  }
\right]}; \hspace{5.1mm} \label{8}
\\
\alpha &=& \frac{1}{\pi} \int_0^\pi {dk\frac{{g - \cos k}}{{t^2
}}{\rm{sech}}^2 \left( \beta \varepsilon_k /2 \right)}; \label{A1}
\\
c_v  &=& \frac{1}{\pi }\int_0^\pi  {dk (\beta \varepsilon_k /2)^2
{\rm{sech}}^2 \left( \beta \varepsilon_k /2  \right)}. \label{S1}
\eea
%In the following we shall analyze and discuss
Armed with these expressions, we now turn to analyzing and discussing the
%their
 crossover and critical scaling behaviors.

\section{Scaling behaviors of thermodynamic quantities in the 1DTFIM} \label{sec:scalingin1DTFIM}
In this section, we discuss the scaling behaviors of
thermodynamic quantities of the 1DTFIM
in the classical disordered region (CDR),
quantum disordered region (QDR), and quantum critical region (QCR)
(Fig.~\ref{fig:1DTFIMphase}).

\subsection{Classical and Quantum disordered regions}
The CDR and QDR in the 1DTFIM are qualitatively characterized by
$\left| {g - 1} \right|/t \gg 1$. In these two regions the free energy density
% follows by
takes the form
\be
f_I \approx  E(g) -\frac{t}{\pi }\int_0^\pi  {e^{ - 2A} dk}  \label{fs}
 \ee
with $A = \sqrt {1 + g^2  - 2g\cos k} /t $,
and $E(g) =  - \frac{1}{\pi }\int_0^\pi  {\sqrt {1 + g^2  - 2g\cos k} dk}$
as its ground state energy density in
% the
 unit of $J$.
The temperature-dependent information is encoded in the integral of Eq.~(\ref{fs})
which can be asymptotically
% solved
performed via the steepest descent method \cite{Wang2000,Lavrentieff2002},
\bea
\frac{t}{\pi }\int_0^\pi  {e^{ - 2A} dk}  \approx \frac{{t^2 }}{{\sqrt {2\pi g} }}f\left( {\frac{{\left| {1 - g} \right|}}{t}} \right) \label{fs2}
\eea
with the universal scaling function $f(\left| {1 - g} \right|/t) = (\left| {1 - g} \right|/t)^{1/2} e^{ - 2\left| {1 - g} \right|/t}$. Therefore in the CDR and QDR, the leading contribution
%for
to the free energy density becomes
\bea
f_I  - E(g) \approx  - \frac{{\left| {1 - g} \right|^2 }}{{\sqrt {2\pi g} }}\left( {\frac{t}{{\left| {1 - g} \right|}}} \right)^{3/2} e^{ - \frac{{2\left| {1 - g} \right|}}{t}}. \label{fs3}
\eea
\begin{figure}[t]
\begin{center}
\includegraphics[width=7.5cm]{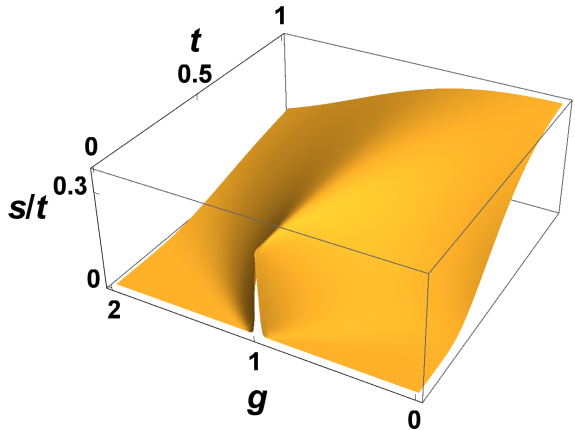}
\end{center}
\caption{
%A 3D plot for the
The entropy (divided by $t$) as a function of
the reduced temperature $t$ and reduced transverse field $g$.
When the temperature is
%large
high the entropy maxima deviate from
the line $g_c=1$, but
when the temperature is
%small
low
the
maxima of entropy appear near the QCP.}
\label{fig:entropy}
\end{figure}
%Obviously
Although Eq.~(\ref{fs3}) includes an exponentially thermal
suppression term,
it still is consistent with the general hyperscaling
ansatz for the free energy in Eq.~(\ref{F0}).
%Then starting from
From Eq.~(\ref{fs3}),
we can readily determine the
critical scaling behaviors of various thermodynamic quantities
%follow as (keep the leading term)
to the leading term as follows
\bea
&& s  \approx \sqrt {\frac{2}{{\pi g}}} \left| {g - 1} \right|^{3/2} t^{ - 1/2} e^{ - \frac{2}{t}\left| {1 - g}
 \right|}; \label{s2} \\
&& c_v  \approx   \sqrt {\frac{{2\pi t}}{{g/\left| {1 - g} \right|}}} \left( {\frac{{\left| {g - 1} \right|}}{t}} \right)^2 e^{ - \frac{2}{t}\left| {1 - g} \right|} ;\label{cv2} \\
&& \alpha \approx   \sqrt {\frac{{2\pi t}}{{g/\left| {1 - g} \right|}}} \frac{{\left| {g - 1} \right|}}{{t^2 }}e^{ - \frac{2}{t}\left| {1 - g} \right|} {\mathop{\rm sgn}} (g - 1); \label{a7} \\
&& \Gamma_{cr}(t \to 0, g)  \approx \frac{\alpha }{{c_v }}  = \frac{{{\mathop{\rm sgn}} (g - 1)}}{{\left| {g - 1}
 \right|}} \label{GQDR}.
\eea
The Gr\"uneisen ratio in Eq.~(\ref{GQDR}) is clearly divergent when approaching
% to
the QCP in the CDR or QDR at low temperatures. A direct
result of Eq.~(\ref{GQDR}) is that at low temperatures the entropy is maximized near
the QCP when the control parameter $g$ is tuned to
%cross
across
the QCP, at $g_c = 1$,
 as discussed in Sec. \ref{sec:general}.
Fig.~\ref{fig:entropy}
%clearly
explicitly shows the
 %maximum behavior
maximization of the entropy near the QCP.
%The results Eqs.~(\ref{s2},\ref{cv2},\ref{a7},\ref{GQDR}) also hold for the
%classical disordered region (Fig.~\ref{fig:1DTFIMphase}).

\subsection{Quantum critical region} \label{sec:1DTFIMinQCR}
When the system stays in the QCR, one immediately obtains
$\Gamma_{cr} (t , g=1) =1/2$,
which
%This result is surprising, because it
deviates from the scaling prediction of Eq.~(\ref{G2}).
%This deviation is understandable because
In order to understand this result, we note that
%for obtaining
 the scaling prediction of Eq.~(\ref{G2})
 % it assumes
has been made based on the assumption\cite{Zhu2003} that
 %the presence of
the linear
% scaling
term in the scaling expansion series of Eq.~(\ref{F1})
in powers of $r/(T/T_0)^{1/(\nu z)}$
is nonzero.
%However, in general this assumption may breakdown, i.e.,
When the linear scaling term
%may vanish while
vanishes, we need to expand the series to the sub-leading
%ones survive,
terms, and this may yield
%leading to
different critical behaviors beyond the prediction of Eq.~(\ref{G2}).
Indeed, from the the scaling form of the free energy in the QCR of the 1DTFIM,
the linear scaling term vanishes and the quadratic one appears
as the leading contribution, resulting in a classical-like constant scaling behavior
for the Gr\"uneisen ratio $\Gamma_{cr}$ near the QCP in the QCR
(the vanishing of $G_T$ in Eq.~(\ref{G2}) for 1DTFIM is also obtained
in Ref.~[\onlinecite{Garst2005}] following a different analysis).

In the QCR, $|1-g| \ll t$ is a natural constraint. We
work in the low-temperature limit.
We note that $A = ((1 - g)^2 /t^2  + 4g\sin ^2 (k/2)/t^2 )^{1/2}$,
and, moreover, there is
a crossover ${k_c} \approx t$ such that when $k < {k_c}$, $\left|
{\sin \left( {k/2} \right)} \right|/t \ll 1$ and when $k > {k_c}$,
$\left| {\sin \left( {k/2} \right)} \right|/t \gg 1$. Correspondingly,
we split the
integration in the expression for the free energy into two parts,
\bea
f_I  = - \frac{t}{\pi } \left( \int_0^{{k_c}} + \int_{{k_c}}^\pi \right) {dk\ln \left(
{{e^{A} + {e^{ - A}}}} \right)}.
\eea
We only consider
the low-lying contributions from $k < {k_c}$, then
$\sqrt
{{{\left( {\frac{{1 - g}}{t}} \right)}^2} + 4g{{\left( {\frac{{\sin
\left( {k/2} \right)}}{t}} \right)}^2}} \ll 1 $. As a result,
\be
 - \frac{t}{\pi }\int_0^{k_c } {dk\ln \left( {e^{A}  + e^{ - A} } \right)}
 \mathop  \approx \limits^{k_c  \approx t}  - \frac{{t^2 }}{\pi
}\left( { a_g + \frac{1}{2}\left(
{\frac{{1 - g}}{t}} \right)^2 } \right),
\ee
where $a_g = \left( {g/6 + \ln 2} \right)$.
It follows that in the low-temperature limit in the QCR,
\be
f_I \sim  - {t^2}\left( {a_g +\frac{1}{2}
\frac{{{{\left( {1 - g} \right)}^2}}}{{{t^2}}}} \right), \label{f2}
\ee
which is consistent with the hyperscaling ansatz. However,
the leading scaling term is quadratic in the expansion parameter $(1-g)/t$.
From Eq.~(\ref{f2}) we obtain
\be
s \sim (2\ln 2 + \frac{{g}}{3})t \Rightarrow \left\{
\begin{array}{l}
 \alpha  \sim t/3 \\
 {c_v} \sim (g/3 + 2\ln 2)t \\
 \end{array}  \right.
 \ee
and ($g$ is fixed near critical point) \be
 \Gamma_{cr}  = \frac{\alpha }{{{c_v}}} \sim {\rm constant}.
\label{ss1}
 \ee
We have thus provided the understanding for the G\"uneisen ratio
reaching a constant in the QCR, $|1-g| \ll t$.  The case we considered earlier,
$g=g_c=1$, falls in this regime. The contradiction with the usual
scaling prediction of Eq.~(\ref{G2}) is only apparent. Generically, in each regime
(such as QCR, CDR or QDR), we can express the scaling function of the free energy
in terms of a scaling variable that is small in that regime [{\it c.f.} Eq.~(\ref{F0})
for the CDR and QDR, and Eq.~(\ref{F1}) for the QCR], and we expect that the leading
linear term in the Taylor expansion is nonzero. Eqs.~(\ref{G2},\ref{G3}) then follows.
%For the 1DTFIM, the QDR belongs to this generic case: for
However, it is special for the 1DTFIM in all regions.
For the CDR or QDR, corresponding to
$t/|g-g_c| \ll 1$,
%the linear term in the Taylor expansion of $\tilde{f}(t/|1-g|)$ is nonzero.
the leading term actually exponentially decays (Eq.~(\ref{fs2})). Nevertheless the
obtained Gr\"uneisen ratio still respects the general analysis Eq.~(\ref{G3}),
though the combined exponents for the
coefficient $G_r$ in Eq.~(\ref{G3}) can not apply to the 1DTFIM.
By contrast, the scaling function in the QCR indeed has a Taylor expansion form.
However, we
have shown that, for the QCR of the 1DTFIM,
the linear term in the Taylor expansion of the scaling function for the free energy,
$f(|1-g|/t)$ vanishes. The leading non-vanishing term is quadratic,
which leads to the apparent violation of Eq.~(\ref{G2}) in the QCR.
(We believe that this mechanism also applies to the anisotropic XY model,
in which a similar constant behavior of the Gr\"uneisen ratio arises in the QCR
\cite{Zhitomirsky2004}.) We stress that our analysis
for the 1DTFIM makes it clear that the the hyperscaling form for the singular
part of the free energy introduced in Ref.~[\onlinecite{Zhu2003}] still applies here.
The constant
behavior of the Gr\"uneisen ratio in the QCR does not violate the scaling form
of the free energy, but is a reflection of a unique form of the scaling function.

\begin{figure}[t!]
\begin{center}
\includegraphics[width=0.8\columnwidth]{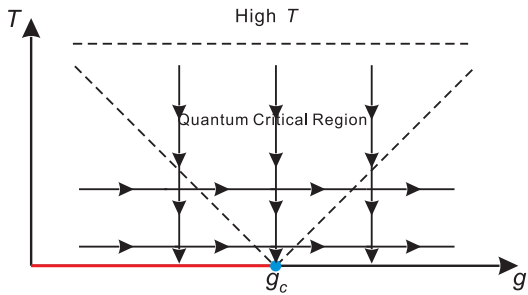}
\end{center}
\caption{The horizontal (vertical) lines of arrows denote
tuning control parameters (temperatures)
across different regions of the 1DTFIM at fixed temperatures
(tuning parameters).} \label{fig:hc}
\end{figure}

\section{Crossover scaling of thermodynamic quantities} \label{sec:crossoverscalingin1DTFIM}

\begin{figure}[t!]
\begin{center}
\includegraphics[width=\columnwidth]{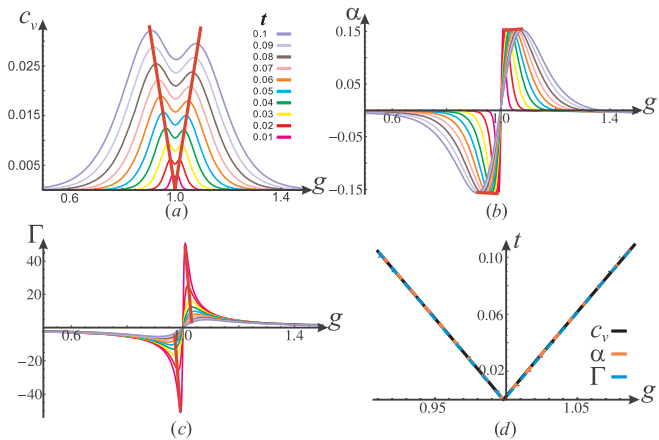}
\end{center}
\caption{The crossover scaling behaviors of the specific heat
[Fig.~\ref{figure:scalingfromg}(a)],magnetic expansion coefficient
[Fig.~\ref{figure:scalingfromg}(b)], the Gr\"uneisen ratio
[Fig.~\ref{figure:scalingfromg}(c)].
%In Figs.~\ref{figure:scalingfromg}(a,c,e) temperatures are fixed from $t=0.01$
%(the curve with peaks nearest to the QCP) to $t=0.1$
%(the curve with peaks furthest from the QCP) with stepwise $t = 0.01$.
The control parameter $g$ varies in the range $(0.5,1.5)$.
The extrema in the figures for each thermodynamic quantity at
each fixed $t$ are illustrated by the red solid line.
Those extrema are then plot on
the $g-t$ plane to identify the corresponding line of the crossover,
which are shown in Fig.~\ref{figure:scalingfromg}(d).
The crossover lines for $c_v$, $\alpha$, and
the Gr\"uneisen ratio can be fitted as $t = 8 \times 10^{-5} + 1.16 | g - 1.00| ^{0.99}$,
$t = 7 \times 10^{-5} + 1.17 | g - 1.00|^{0.99}$, and $t = 8 \times 10^{-5} + 1.18 | g - 1.00 | ^{0.99}$, respectively.
The exponent 0.99 very close to the exact scaling exponent 1.
The crossover lines almost overlap with each other.}
\label{figure:scalingfromg}
\end{figure}
Based on different dominant factors, the phase diagram
of the 1DTFIM can be qualitatively divided into three regions: the low-temperature
CDR, QDR, and the QCR, as shown in Fig.~\ref{fig:1DTFIMphase}.
Near the boundaries of the different regions, there is a strong
competition between the thermal and quantum fluctuations and, as a result,
crossover signatures are expected in physical properties
as we tune the transverse field or temperature across the boundaries.
In the following we shall demonstrate that there indeed exist crossover scaling
behaviors in the 1DTFIM for the specific heat and magnetic expansion coefficient.
However, subtleties emerge for the Gr\"uneisen ratio in the QCR.

\subsection{Crossover at Fixed Temperatures} \label{sec:scalingvsg}
Consider first the isothermal behavior as illustrated in
Fig.~\ref{fig:hc}, where each horizontal line of arrows illustrates
tuning the transverse magnetic field through the different
regions of the phase diagram at a fixed (low) temperature $t$.
The results are plotted in Figs.~\ref{figure:scalingfromg}(a,b,c).
They show that the thermodynamic quantities of the 1DTFIM reach extrema
when $g$ is tuned across the boundaries of the different regions.
The crossover scaling exponent extracted from the numerical results
of Figs.~\ref{figure:scalingfromg}(d) is $0.99$
which agrees well with the universal scaling exponent in the 1DTFIM,
$z \nu = 1$.
The minor deviation of the exponent from the exact value
is because when the temperature is relatively high the
thermal fluctuations become stronger such that
those crossover signatures in the crossover are weakened,
making it harder to precisely locate the crossover. However, the deviation is negligible,
indicating the existence of a large region of quantum criticality in the 1DTFIM,
which is consistent with the conclusions drawn from a recent NMR experiment
on the quasi-1D ferromagnet CoNb$_2$O$_6$ \cite{Kinross2014}.

\subsection{Crossovers at Fixed Control Parameters}
%\begin{figure}[t]
%\begin{center}
%\includegraphics[width=0.8\columnwidth]{vc.eps}
%\end{center}
%\caption{The lines of arrows illustrate tuning temperatures across different regions of the 1DTFIM at fixed
%control parameters.} \label{fig:vc}
%\end{figure}

\begin{figure}[t]
\begin{center}
\includegraphics[width=\columnwidth]{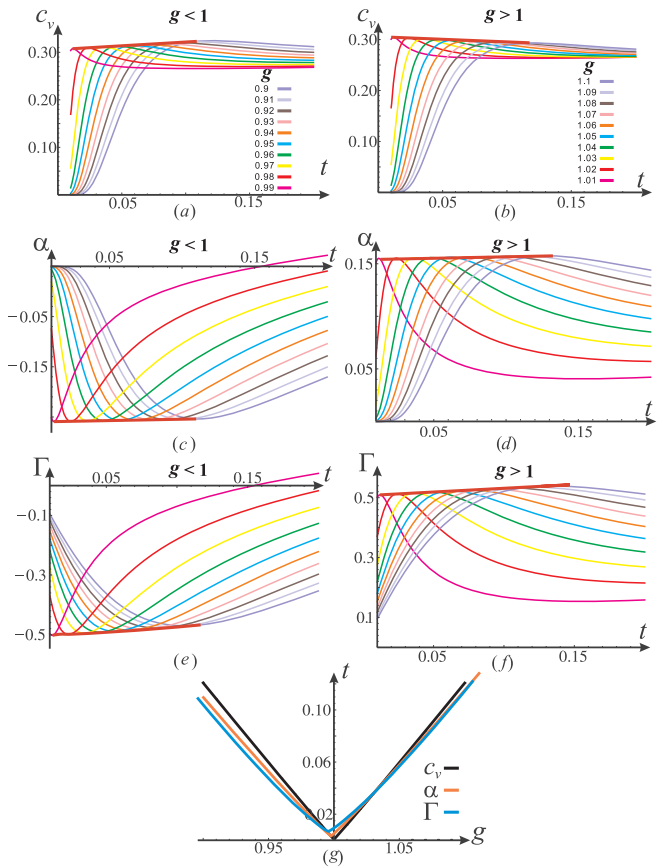}
\end{center}
\caption{The crossover scaling behaviors of the specific heat
[Figs.~\ref{figure:scalingfromt}(a,b)], magnetic expansion coefficient
[Figs.~\ref{figure:scalingfromt}(c,d)], and Gr\"uneisen ratio
[Figs.~\ref{figure:scalingfromt}(e,f)].
%In Figs.~\ref{figure:scalingfromt}(a,d,g)
%[Figs.~\ref{figure:scalingfromt}(b,e,h)] the tuning parameters are fixed from
%$g=0.9$[$g = 1.1$] (the curve with the peak furthest from $t=0$)
%to $g=0.99$ [$g=1.01$] (the curve with the peak nearest to $t$=0)
%with stepwise $g = 0.01$.
The temperature is set in the range $(0.01,0.2)$.
The extrema in the figures for each thermodynamic quantity at
each fixed $g$ are illustrated by the red solid line.
Those extrema are then plot on $g-t$ plane to fit the crossover scaling behaviors,
shown in Fig.~\ref{figure:scalingfromt}(g).
The crossover scaling for $c_v$, $\alpha$,
and $\Gamma$ fit as $t = 6 \times 10^{-4} + 1.27 | g - 1.00| ^{1.02}$,
$t = 0.003 + 1.38 | g - 1.00|^{1.07}$, and $t = 0.007 + 1.73 | g - 0.99 | ^{1.18}$,
respectively. } \label{figure:scalingfromt}
\end{figure}
The crossovers at fixed $g$ are illustrated in Fig.~\ref{fig:hc},
where each vertical line of arrows represents temperatures tuning across
different regions at a fixed tuning parameter $g$.
Similarly, we use the extrema in Figs.~\ref{figure:scalingfromt}(a,b,c,d,e,f)
to extract the crossover of $c_v$, $\alpha$ and the Gr\"uneisen ratio.
Fig.~\ref{figure:scalingfromt}(g) shows that
the crossover scaling of the specific heat coefficient
and magnetic expansion coefficient are consistent
with the scaling prediction with scaling exponent $\nu z =1$.
It is also coincident with previous discussions
on the crossover scalings with tuning control parameters at fixed temperatures.
However, for the Gr\"uneisen ratio, its crossover scaling exponent
significantly deviates from the general crossover scaling argument,
as it is shown in Fig.~\ref{figure:scalingfromt}(g).
The crossover scaling with power-law fitting gives rise to an exponent of $1.18$,
which is about $20\%$ larger than the exact value! This reflects the
weaker form of the critical singularity in the Gr\"uneisen ratio
as a function of temperature, as we discussed in Sec.~\ref{sec:1DTFIMinQCR}
and seen in Fig.~\ref{figure:scalingfromt}(g).

%And when $g$ approaches to the QCP,
%the scaling behavior of the Gr\"uneisen ratio approaches to a constant
%instead of collapsing on the crossover scaling function $t \sim |1-g|$,
%different from the scaling behaviors of other thermodynamic quantities.
%This is consistent with general discussions in Sec.~\ref{sec:1DTFIMinQCR}.

\section{Discussion and Conclusions} \label{sec:discussions}

In this paper, we have studied in some detail the scaling behavior
of thermodynamic properties in the 1DTFIM at low temperatures.
The results are unusual and can serve as telltale signs of the 1DTFIM:
the Gr\"uneisen ratio is power-law divergent following the conventional
scaling analysis when the system is tuned across the QCP as a function
of the non-thermal control parameter (the transverse magnetic field) at
zero temperature; by contrast, it approaches a constant when the QCP is
approached with a decreasing temperature in the quantum critical region.

We clarified the reasons for this unusual feature. The singular part of
the free energy satisfies the form introduced earlier (Ref.~[\onlinecite{Zhu2003}]).
However, its scaling function is unique in that, in the quantum critical
region (but not in the quantum disordered or classical disordered region),
the scaling function can be Taylor expanded and its linear term vanishes.
This unsual form of the scaling function makes the temperature dependence
of the Gr\"uneisen ratio in the quantum critical region to differ from the
generic expectations of the scaling analysis. In the quantum disordered and
classical disordered regions,
%the scaling functions have the generic form
%in that the linear terms in the corresponding Taylor expansions are nonzero.
%This ensures the power-law divergent form of the Gr\"uneisen ratio as the
%non-thermal control parameter is tuned to the quantum critical value in these
%two regimes.
the leading behavior of the scaling functions is dominated by a peculiar
exponential form different from the usual Taylor expansion one. Despite of
lacking simple combined exponents for $G_r$ (Eq.~(\ref{G3})) used in Ref.~[\onlinecite{Zhu2003}],
the obtained power-law divergent form of the Gr\"uneisen ratio as
a function of
the non-thermal control parameter still respects the results
obtained
%from conventional Taylor expansion analysis
in these two regimes.
Consequently, entropy is enhanced as the transverse field is
tuned to its critical value at low but nonzero temperatures,
as expected from generic scaling analysis.

We have also discussed the crossover behavior of the specific heat,
magnetic expansion coefficient and the Gr\"uneisen rato. The contrast
between the critical behavior of the Gr\"uneisen ratio between the
quantum critical and quantum or classical disordered regions is also manifested
in how the thermodynamic quantities capture the crossovers when they
are approached from different directions in the phase diagram of
temperature and transverse magnetic field.

The one-dimensional transverse-field Ising model is a paradigmatic
theoretical model for quantum criticality, and our work clarifies
the singularities, scaling and crossover of the thermodynamic quantities
in this model. Given the wide interest in quantum criticality,
it would be important to have actual materials that realize the
one-dimensional transverse-field Ising model. At the present time,
the materials which are well-established to be describable by the
one-dimensional transverse-field Ising model is still rare. The unique
forms of critical scaling and crossover we have determined in this work
will serve as signatures to experimentally identify materials that
realize this prototype model system for quantum criticality.
Finally,
the one-dimensional transverse-field Ising model has a fermionic representation
[Eqs.~(\ref{3}, \ref{6})], ergo, our results
can also shed light on understanding the critical scaling behaviors in
the relevant fermionic model.

\emph{Note added}: After this work was completed, we learnt the experimental
studies in a  quasi-one-dimensional material had provided evidences to
support our scaling predictions [\onlinecite{Wang2018}].

%we learnt of experimental
%studies in a  quasi-one-dimensional material that is suitable to test
%our scaling predictions [Z. Wang et al., private communications (2018)].

\section{Acknowledgment}

We would like to acknowledge useful discussions
with Zhe Wang and Robert-Jan Slager. This work  was in part supported by
the NSF grant No.\ DMR-1611392 and the Robert A. Welch Foundation
Grant No.\ C-1411. Q.S. acknowledges the hospitality of the
Aspen Center for Physics (the NSF Grant No. PHY-1607611).
Two of us (L.Z. and Q.S.) would like to thank M. Garst and A. Rosch
for an earlier collaboration (Ref.~[\onlinecite{Zhu2003}]) on this general topic.

\bibliography{TFIM}
\bibliographystyle{apsrev-nourl}

\end{document}